\documentclass{amsart}

\usepackage{graphicx}
\usepackage{url}

\newcommand{\bq}{\mathbf{q}}
\newcommand{\bx}{\mathbf{x}}
\newcommand{\bv}{\mathbf{v}}
\newcommand{\bg}{\mathbf{g}}
\newcommand{\ba}{\mathbf{a}}
\newcommand{\bb}{\mathbf{b}}
\newcommand{\bc}{\mathbf{c}}
\newcommand{\bfh}{\hat{\mathbf{f}}}
\newcommand{\bpsi}{\mathbf{\Psi}}

\newcommand{\mL}{\mathcal{L}}
\newcommand{\mE}{\mathcal{E}}
\newcommand{\coslaw}{{\rm coslaw}}

\theoremstyle{definition}

\theoremstyle{remark}

\usepackage[margin=1in]{geometry}

\begin{document}

\title[Cosmological Origami]{Cosmological Origami: Properties of Cosmic-Web Components When a Non-Stretchy Dark-Matter Sheet Folds}
\def\publname{Submitted, Proceedings of the 6th International Meeting on Origami in Science, Mathematics and Education}

\author{Mark C. Neyrinck}
\address{Department of Physics and Astronomy, The Johns Hopkins University, Baltimore, MD 21218, USA}
\email{neyrinck@pha.jhu.edu}
\thanks{I am grateful for support from a New Frontiers in Astronomy and Cosmology grant from the John Templeton Foundation, and from a grant in Data-Intensive Science from the Gordon and Betty Moore and Alfred P. Sloan Foundations.}



\begin{abstract}
In the current cosmological paradigm, an initially flat three-dimensional manifold that pervades space (the `dark-matter sheet') folds up to build concentrations of mass (galaxies), and a cosmic web between them. Galaxies are nodes, connected by a network of filaments and walls.  The folding is in six-dimensional (3D position, plus 3D velocity) phase space. The positions of creases, or caustics, mark the edges of structures.\\
\\
Here, I introduce an origami approximation to cosmological structure formation, in which the dark-matter sheet is not allowed to stretch. But it still produces an idealized cosmic web, with nodes, filaments, walls and voids. In 2D, nodes form in `polygonal collapse' (a twist-fold in origami), necessarily generating filaments simultaneously. In 3D, nodes form in `polyhedral collapse,' simultaneously generating filaments and walls. The masses, spatial arrangement, and angular momenta of nodes and filaments are related in the model. I describe some `tetrahedral collapse', or tetrahedral twist-fold, models.
\end{abstract}

\maketitle

\section{Introduction}
\begin{figure}
    \begin{center}
     \includegraphics[width=\columnwidth]{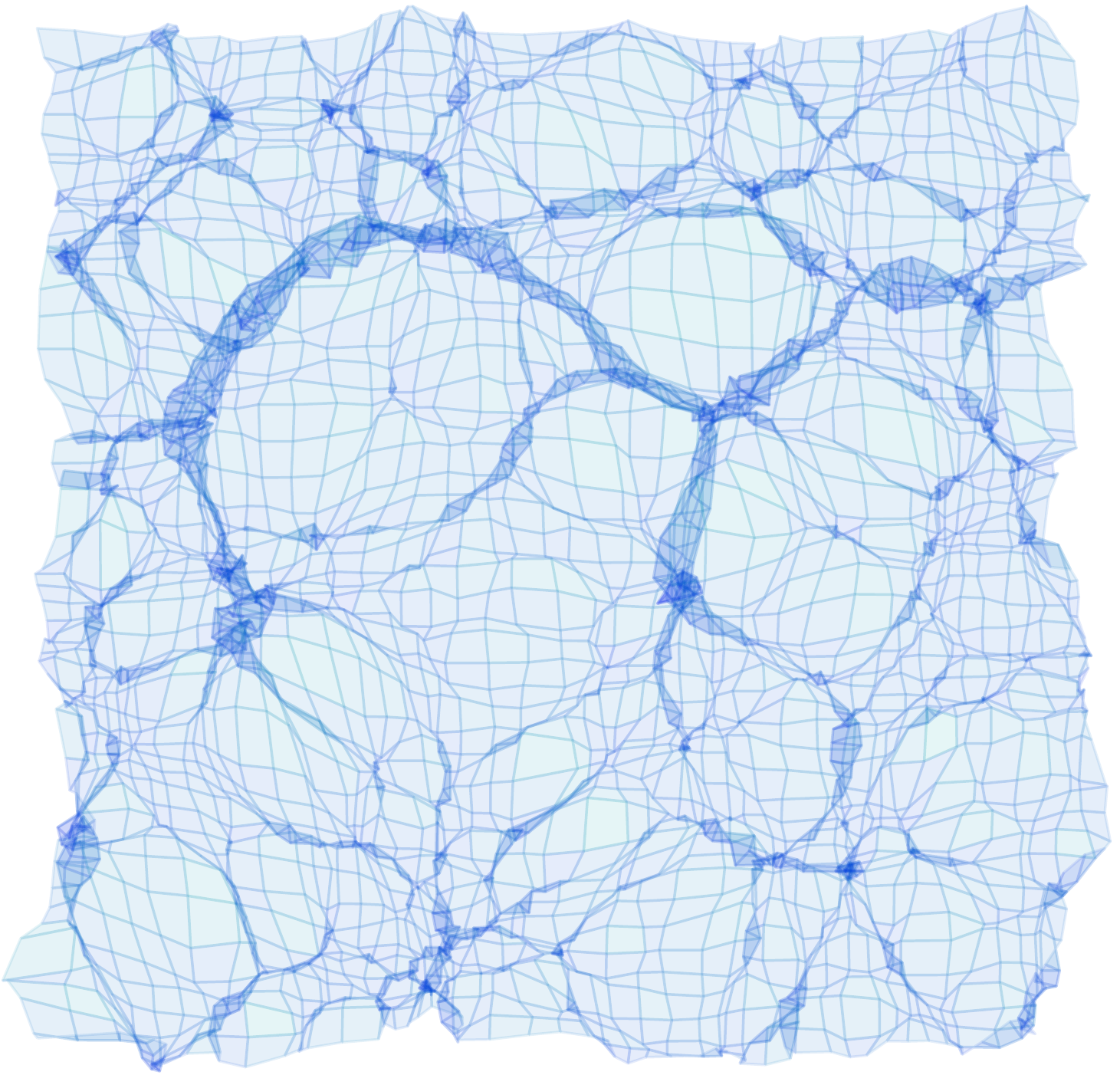}
    \end{center}  
   \caption{A dark-matter sheet in a 2D universe, that distorts and folds through an approximation to gravity, called the Zel'dovich approximation. The darkness of the color at each position gives the number of streams there. Initially, all vertices were nearly on a regular lattice. Since then, gravity has distorted the mesh, causing regions with a bit more matter than average to accumulate more matter around them. The patch shown is $>10^8$ light-years on a side; nodes correspond to galaxies or clusters of galaxies.}
    	\label{fig:fyou}
\end{figure}

The formation of structure in the Universe proceeds somewhat like the origami-folding of a sheet. This concept is a Lagrangian fluid-dynamics framework (following the mass elements, not staying in a fixed spatial coordinate system). This approach in cosmology started with the Zel'dovich approximation \cite{Zeldovich1970}. Catastrophe theory has given some further understanding into the types of singularities that can occur when this sheet begins to fold \cite{ArnoldEtAl1982,HiddingEtal2014}. Recently, many have realized the power of Lagrangian dynamics in general, and of explicitly following the dynamics of the sheet in a cosmological simulation, instead of considering the particles within it to be just fuzzy, isotropic blobs of matter \cite{ShandarinEtal2012,AbelEtal2012,FalckEtal2012}. Fig.\ \ref{fig:fyou} shows an example cosmic web \cite{BondKofmanPogosyan} folded from a collisionless dark-matter sheet that has distorted and moved around according to the Zel'dovich approximation.

\section{An Origami Approximation to Structure Formation}
Paper flat-origami is conveniently expressed as a continuous, piecewise isometry defined on a finite region, say $[0,1]^2\to\mathbb{R}^2$, encodable as a set of reflections (creases). Cosmological origami has similarities: it can be thought of as a mapping from a 3D (`dark-matter') sheet, that can cross itself without resistance in 3D (since dark matter is thought to be collisionless), but cannot cross itself in 6D position-velocity phase space.  The dark-matter sheet, and the universe in which it folds, are (as far as we can tell) infinite. To avoid dealing with an actual infinite space, often a finite domain with periodic boundary conditions (toroidal topology) is considered in cosmology, for example in $N$-body gravitational simulations.  To investigate how nodes interact with each other as they fold into a network, it might be interesting to look at origami in a region with periodic boundaries. But for this introductory exploration, we concentrate on properties of isolated structures in an infinite space.

Some definitions: the analogy of a `crease pattern' in cosmology is called {\it Lagrangian space}, abbreviated $\mL$. The {\it Lagrangian position} $\bq\in\mL$ of a particle specifies its initial position on the `crease pattern.' The usual position space is called {\it Eulerian space}, $\mE$. After gravity folds $\mL$ up into structures, the actual position of a particle at time $t$ is called its {\it Eulerian position}, $\bx(\bq,t)\in\mE$. All coordinates here are comoving, meaning that the homogeneous, isotropic expansion of the universe is scaled out; one can think of this as zooming out at the same time as the expansion happens. Initially, in $\mL$, particles were arranged almost uniformly. They had vanishing velocity as $t\to0$, so initially, in position-velocity phase space (3D position, plus 3D velocity), they were arranged on a flat sheet, flat (everywhere zero) as viewed in the velocity dimensions.

The {\it displacement field} $\bpsi$ gives the vector between the folded-up and initial positions of a particle, $\bpsi(\bq,t)\equiv\bx(\bq,t)-\bq$, and the velocity is just the time derivative of the Eulerian position, $\bv=\partial \bx/\partial t$. A {\it stream} is a region of $\mL$ delimited by caustics, or folds \cite{Neyrinck2012}. In an origami crease pattern, this would correspond to a polygon surrounded by creases.  A {\it void} is a single-stream region, i.e.\ a region in which the mapping from initial to final position, $\bx(\bq)$, is one-to-one \cite{Shandarin2011,FalckEtal2012}.

The origami approximation imposes the following assumptions on the functions $\bpsi(\bq)$ or $\bx(\bq)$.  `Reality' means `the current cosmological structure-formation paradigm,' which fits many observations quite well.
\begin{enumerate}
\item $\psi(\bq,t)\equiv|\bpsi(\bq,t)|$ is bounded over all $\bq\in\mathbb{R}^3$. This property holds in reality; the distribution of components $\Psi_i$ over all particles is a roughly Gaussian distribution of dispersion $\sim 5\times10^7$ light-years, small compared to the radius of the observable universe, $\sim5\times10^{10}$ light-years.
\item $\bpsi$ is irrotational in a void. This holds in reality, since any initial vorticity decays with the expansion of the Universe, and since gravity is a potential force. However, in {\it multistream} regions (where $\bx(\bq)$ is many-to-one), the flow, averaged among streams, often carries vorticity.
\item The dark-matter sheet does not stretch, for a close analogy to paper origami. That is, $\bx(\bq)$ is continuous and piecewise-isometric, i.e.\ $\mathbf{\nabla}_{\rm Lagrangian}\cdot\bpsi=0$, except at creases, where it is undefined.  This is the only assumption that is manifestly broken in reality, but we explore the consequences which it helps to establish, which we hope will apply more generally.
\end{enumerate}

\begin{figure}
  \begin{minipage}[b]{0.35\linewidth}
    \begin{center}
     \includegraphics[width=1.2\columnwidth]{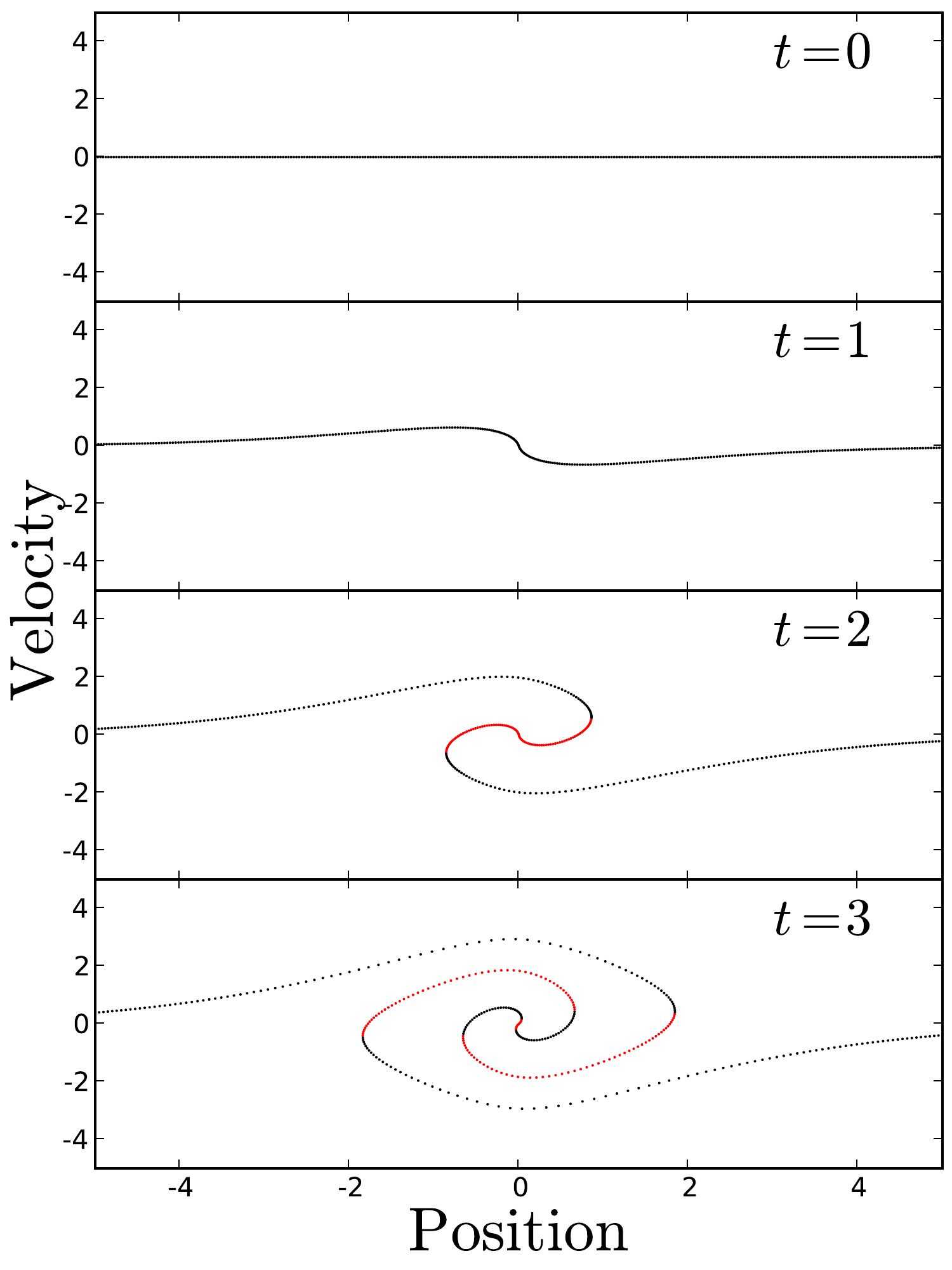}
    \end{center}  
  \end{minipage}
  \begin{minipage}[b]{0.64\linewidth} 
    \begin{center}
      \leavevmode
    \end{center}
  \caption{A schematic phase-space spiral that occurs in a 1D collapse of the dark-matter `string' (sheet in $>1$D), on which vertices are represented as dots. Caustics, or creases, occur where the string goes vertical. Black segments are oriented the same way as at $t=0$, while red segments are oriented oppositely. Note that here, the dark-matter string stretches, i.e.\ the particles vary in their distances. But, the same pattern of creases can be produced without stretching.}
    	\label{fig:spirals}
    \end{minipage}
\end{figure}

Collapsed structures in 1D are simple. In 1D, a {\it node} is a connected region of $\mathcal{N}\in\mL$, such that $\bx(\mathcal{N})$ is multistream, but points immediately outside $\mathcal{N}$ map to voids.   Fig.\ \ref{fig:spirals} shows a schematic, but realistic, node in a 1D universe. For simplicity, hereafter we will restrict our attention to nodes without substructure. In 1D, nodes are simple pairs of crease points in $\mL$, as shown at $t=2$. We disregard added creases (e.g.\ present at $t=3$) that leave the extent of the node after folding unchanged.

\subsection{2D: Filaments and Polygonal-Collapse Nodes} 
A 2D universe is useful to investigate; it is a step to 3D, and it also has some relevance in a 3D universe, where galaxies often form within the effective 2D universe of a `wall' (defined below). Under our assumptions, the structure foldable with the fewest creases in 2D is a {\it filament}, i.e.\ a pleat, or pair of parallel crease lines.  A one-crease structure is disallowed since any reflection of half the space would give an unbounded $\psi$. Why must a filament consist of parallel creases? Any crease must be a straight line in a piecewise-isometry \cite{DemaineOrourke2008}.  And, the creases must be parallel, by both assumptions 1 and 2: (1) would be violated if non-parallel creases diverged arbitrarily far at infinity, giving an unbounded $\psi$. More importantly, (2) restricts even finite filaments to consist of parallel creases. This is because the two reflections produced by non-parallel creases would cause neighboring voids to be rotated with respect to each other.

Now consider a 2D node, a finite connected region that is collapsing (becoming multistream after folding). Note that  circular (spherical) collapse is impossible with isometry, because creases must be straight lines, implying that a `collapsing' (becoming multistream after folding) patch must be polygonal. Circular (spherical) collapse can be seen as a limiting case, with isotropy around the node, and in which the sheet stretches substantially.  Polygonal (polyhedral) collapse can be seen as another limiting case, with anisotropy, but no stretching.

By Kawasaki's theorem \cite{Kawasaki1989creases} (the alternating sums of vertex angles in a flat-folding 2D crease pattern add to $180^\circ$), any vertex in a 2D crease pattern must join an even number ($\ge 4$) of creases. Thus, the node cannot form in complete isolation; other structures, e.g.\ filaments, must form together with it. Here, we confine attention to simple nodes in 2D with 4 vertices. Note that Kawasaki's theorem applies to a non-flat manifold, as well, if angles are measured arbitrarily close to a vertex \cite{Robertson1978}.

\begin{figure}
    \begin{center}
     \includegraphics[width=\columnwidth]{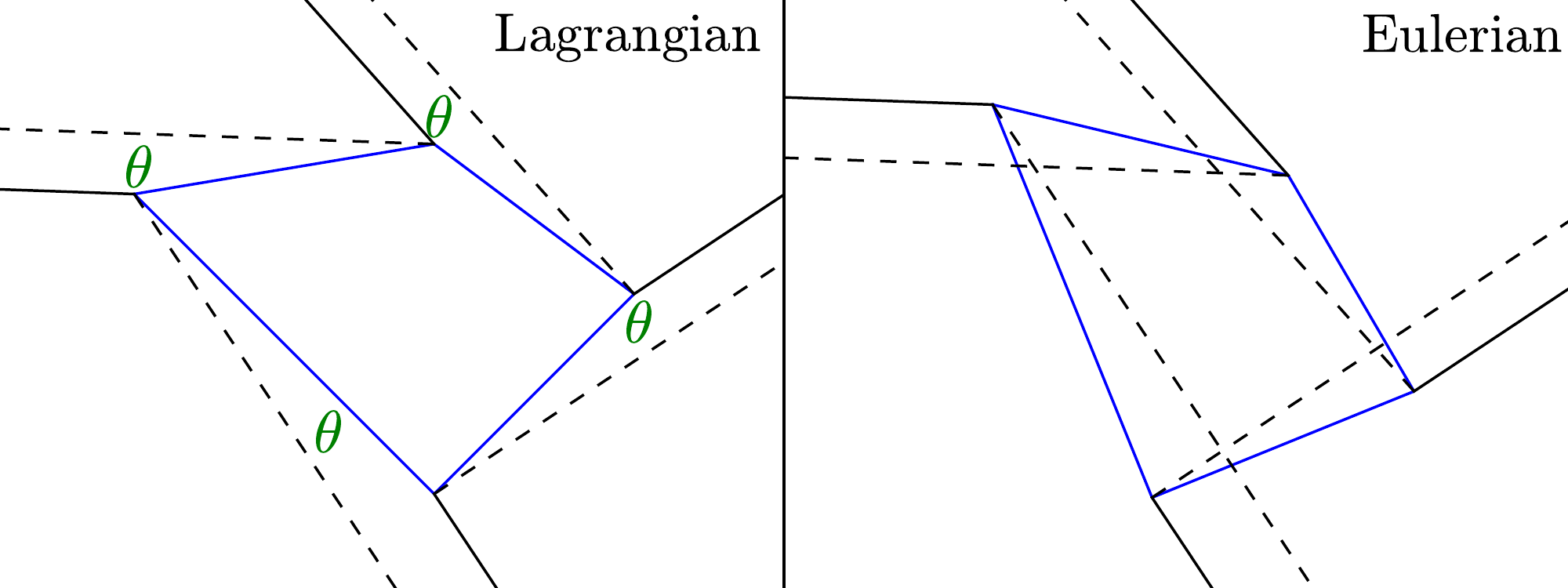}
    \end{center}  
   \caption{A node produced from a clockwise-twisting quadrilateral, before (left) and after (right) folding. Solid/dashed lines show mountain/valley folds, for a paper model. Mountain/valley crease assignment may not be meaningful in 4D phase space.}
    	\label{fig:quadfold}
\end{figure}

Angles at which filaments come off of the node's edges must equal each other (e.g.\ \cite{Kawasaki1997}). Why? By Kawasaki's theorem, the angle opposite $\theta$ at a vertex in Fig.\ \ref{fig:quadfold} is $180^\circ-\theta$. This angle is also opposite an angle in the next (going counterclockwise) vertex in a pair of parallel lines, so this angle must also equal $\theta$. Continuing around the polygon, all angles labeled $\theta$ must be equal.

We have not yet discussed mountain or valley folds, or sheet-crossing. This is because these differ from the case of usual paper origami, in which folding occurs in 3D.  Here, folding is in 4D symplectic phase space, where it is not clear that mountain/valley fold assignment is necessary, and the extra dimension makes sheet-crossing in 4D much more rare. To address possible sheet-crossing, and also to keep a close relation to physical flows, it is useful to adopt a kinematic flat-origami model, in which creases change continuously with time from an initial creaseless state. In polygonal collapse, if the shape of the polygon is held fixed, two parameters may be tuned: $\theta$, as in Fig.\ \ref{fig:quadfold}, and the scale (e.g.\ longest diameter) of the polygon. Either or both of these may be continuously changed from zero to yield a kinematic model. Alternatively, the shape of the polygon can be changed. We have determined velocity fields from many simple models ch.pdfanging either $\theta$ or the scale. In each case, streams overlapping in position have had different velocities, satisfying the no-sheet-crossing condition in 6D phase space. Examples are given in \cite{Neyrinck2014}  The question of when the sheet-crossing condition might be violated in a kinematic model is an obvious question to answer more rigorously, though. Note that, a bit unphysically, velocity fields in such kinematic models can be discontinuous spatially at caustics in $\mL$, and also discontinuous in time, if the caustic is moving.

\begin{figure}
    \begin{center}
     \includegraphics[width=\columnwidth]{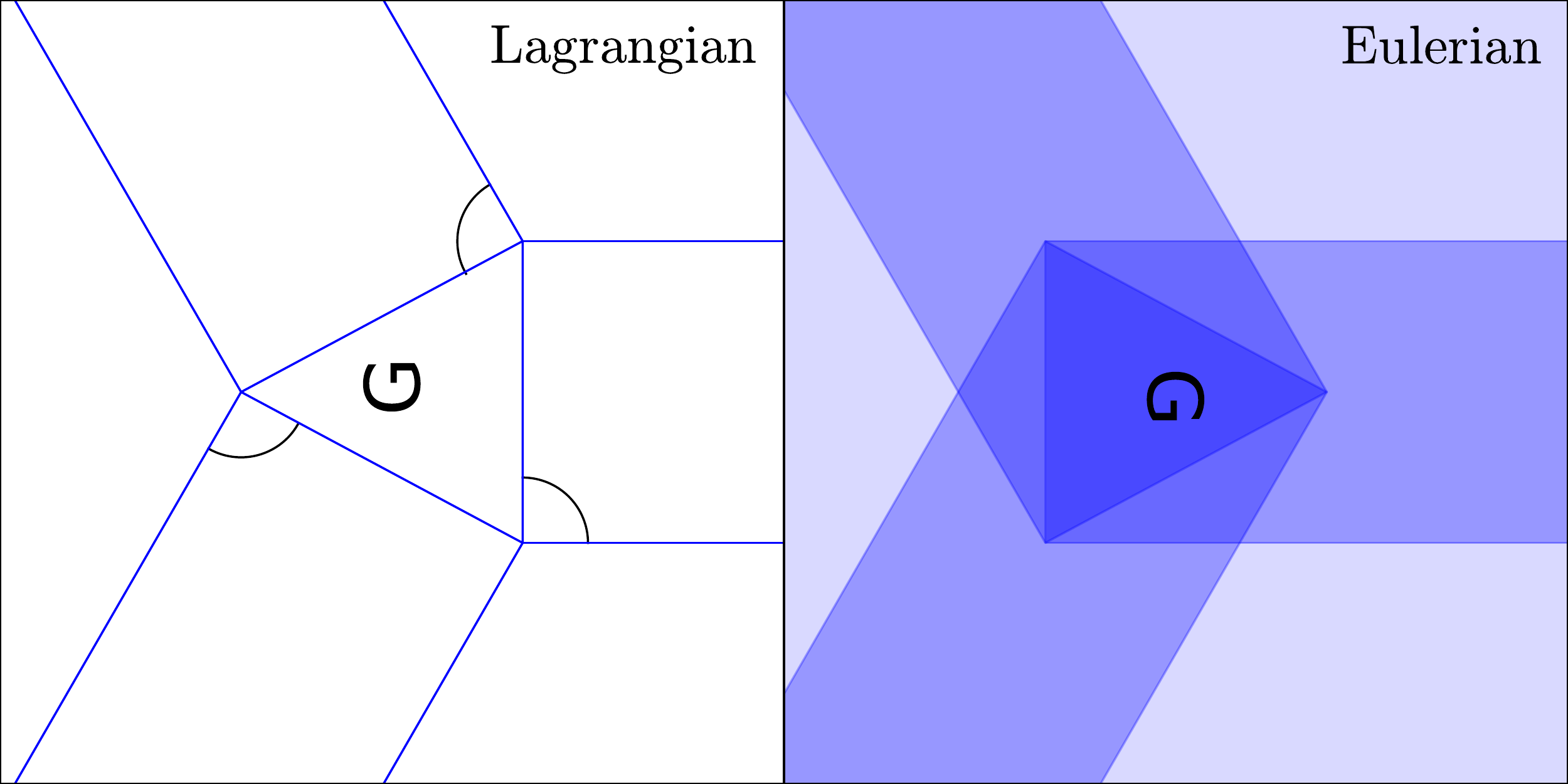}
    \end{center}  
   \caption{Irrotational triangular collapse. Eulerian regions exist with 1, 3, 5, and 7 (at the center) streams.}
    	\label{fig:foldedgal}
\end{figure}

Any $\theta<90^\circ$ rotates the central polygon by $2\theta$; this provides a heuristic way to understand the prevalence of both rotating galaxies and filaments radiating from them. (This can also be seen as a rotation in the other direction by the larger, obtuse angle in the filament, by an angle $2(180^\circ-\theta)=-2\theta$ in the opposite direction.) But in the special case of $\theta=90^\circ$, shown in Fig.\ \ref{fig:foldedgal}, the collapse is an irrotational parity inversion. So, an interesting irrotational model keeps $\theta=90^\circ$, and increases the collapsing polygon's scale from 0. Irrotational collapse can occur with an arbitrarily-many-sided polygon; however, it is unlikely that an arbitrarily complicated such design can be folded from paper.  Already, Fig.\ \ref{fig:foldedgal} seems unable (based on failed attempts) to be folded with paper in a typical twist-fold manner, in which edges of the central polygon are mountain-folded, and the polygon ends up on top. But it can be folded up if one of the triangle's edges is valley-folded. After folding, all streams in the initial crease pattern overlap at the center, giving 7 streams.

Another special case of polygonal collapse is worth mentioning: the intersection of filaments, forming sequentially. If two perpendicular filaments form sequentially, the result is simply a rectangular collapse, a four-sided version of Fig.\ \ref{fig:foldedgal}. If they are not perpendicular, though, the central parallelogram undergoes a rotation. Intersections of more than two sequentially-forming filaments do not generally produce simple nodes with 4-crease vertices. It would be interesting to investigate the properties of 2D nodes, too, that fold on top of each other.

What are the topologies and shapes of voids and multistream regions in a piecewise-isometric origami cosmic web? Since filaments must consist of straight, parallel lines, they are either infinite, or terminate at nodes. Thus, all voids are entirely enclosed by polygons (or are infinite); they form origami tessellations \cite{Gjerde2008}. Angles between filaments are $<180^\circ$, since nodes consist of convex polygons, each vertex of which sprouts a filament. Thus, voids, too, must be convex in the origami approximation.

\subsection{3D: Walls, Polygonal-Collapse Filaments, and Polyhedral-Collapse Nodes}
Origami-folding a 3D sheet is much-less-studied than a 2D sheet. But the two-coloring of streams (regions bounded by creases) in $\mL$ still applies \cite{Neyrinck2012}. In 2D, by the four-color theorem, the number of colors necessary to color arbitrary (non-origami) regions is 4; in 3D, it is unbounded \cite{Guthrie1880}. So the two-colorability condition for streams in $\mL$ is much more restrictive in 3D, reducing the number of colors from $\infty$ to 2, rather than from 4 to 2.

In 3D, a {\it wall} is a parallel pair of creases (plane segments, in 3D). In the same way as in the 2D-universe filament, a wall's creases are constrained to be parallel by voids' mutual irrotationality. In 3D, a {\it filament} can be constructed by extruding a 2D polygonal node along an axis out of its plane (not necessarily $\perp$ to the plane). As in 2D, to maintain mutual irrotationality between voids, the lines extruded from vertices of a 2D node that becomes a 3D filament must remain parallel. A 3D {\it node} lies at the intersection of filaments, and consists of a convex polyhedron enclosed by creases. The 3D generalization of Kawasaki's theorem, discussed below, ensures convexity. As in 2D, 3D filaments and nodes cannot form in isolation; nodes sprout filaments and walls, and filaments sprout walls.

As in 2D sequential-filament-intersections, a simple way to construct a node in 3D is by sequentially folding 3 intersecting walls. Three perpendicular, equal-width walls would give cubic collapse. A bit more complicated model involves a pair of non-orthogonal walls, collapsing along with a third wall, perpendicular to both.  This is essentially a 2D collapse, occurring within the plane of the third wall, and  would impart some rotation to the inner parallelepiped node. An intersection of three arbitrarily oriented walls would also be interesting to investigate.

However, an intersection of three walls produces a node terminating 6 filaments, and 12 walls (double-counting pairs of collinear filaments, and quadruple-counting coplanar walls). This far exceeds the number of structures from nodes typically in a cosmological simulation; there are often as few as 3 filaments emerging from galaxies \cite{DekelEtal2009,DanovichEtal2012}. This suggests that many nodes form within previously-formed walls, and also motivates the following study of a minimal 3D collapse.

\subsection{Tetrahedral Collapse: A Tetrahedral Twist-Fold}
Imagine a point, with four rays coming from it, that do not all point into the same half-space. This breaks space into four void regions, between triplets of rays. In tetrahedral collapse, the voids push together. Because the continuity of the displacement field $\bpsi$ must be preserved, a node (at the point), 4 filaments (along the rays), and 6 walls (between pairs of rays) arise as this happens.

We set up the problem by fixing the vertices of the central tetrahedral node in $\mL$, and constrain the properties of the filaments coming off of it. Denote the vertices of the tetrahedral node $\bg_i$, for $i\in{1,2,3,4}$. Set $\bg_1=(0,0,0)$. This leaves 9 degrees of freedom (DoFs), which can alternatively be assigned to one scale parameter, five parameters for the shape \cite{ShapeDoF}, and three angles that describe a three-dimensional rotation. The rotational DoFs control the relative rotation between the set of filaments and the node.

Filament 1 is an extrusion along direction $\bfh_1$ of the triangle with vertices $\bg_{2,3,4}$, and similarly for filaments $2$-$4$. We set $\bfh_1=(0,0,1)$ along the $z$-axis, and $\bfh_i=(\sin\theta_i\cos\phi_i, \sin\theta_i\sin\phi_i, \cos\theta_i)$ for $i=2$-4, with $\phi_2=0$.

The problem reduces to finding five angles, given all $\bg_i$. There are two meaningful sets of constraints. First, the angles at which walls come off of each filament $i$ must equal each other; this is the triangular-collapse constraint applied in the plane $\perp \bfh_i$. Second, the creases coming from vertices must obey the 3D generalization of Kawasaki's theorem (Kawasaki-3D) \cite{Robertson1978,
Kawasaki1989,Hull2010}: two-color the corners of regions that meet in a vertex, such that one color represents original parity when folded, and the other color represents opposite parity. The sum of solid angles of each color equals $2\pi$ steradians. In tetrahedral collapse, voids, walls, filaments, and the node have parities 1, -1, 1, and -1, respectively. 

\begin{figure}
  \begin{minipage}[b]{0.3\linewidth}
    \begin{center}
     \includegraphics[width=1.2\columnwidth]{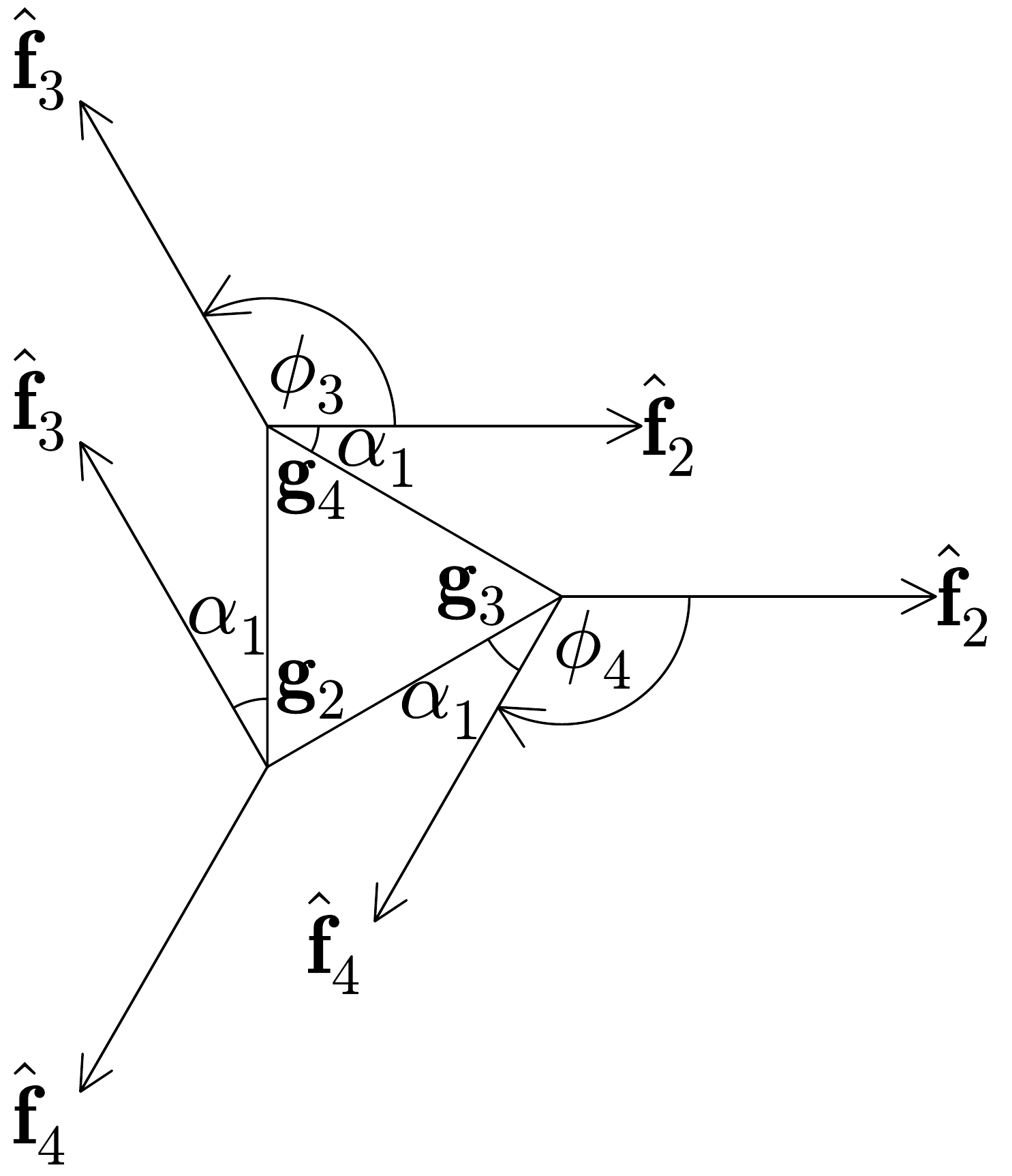}
    \end{center}  
  \end{minipage}
  \begin{minipage}[b]{0.69\linewidth} 
    \begin{center}
      \leavevmode
    \end{center}
  \caption{A cross-section through filament 1, with $\bfh_1\perp$ the page. All vectors should in fact have a `$_{\perp 1}$' subscript, denoting the component of the vector perpendicular to $\bfh_1$. Because of our choice of axes ($\bfh_1=\hat{z}$ and $\phi_2=0$), $\phi_3$ and $\phi_4$ come out simply, looking down the barrel of filament 1.}
    	\label{fig:filamentbarrel}
    \end{minipage}
\end{figure}

Looking down the barrel of a filament, in the plane perpendicular to each $\bfh_i$, triangular collapse must happen, with walls coming off of vertices at equal angles. Considering filament 1, shown in Fig.\ \ref{fig:filamentbarrel}, $\phi_4$ and $\phi_3$ come out easily from vectors involving $\bg_{2\perp1}$, $\bg_{3\perp1}$, and $\bg_{4\perp1}$, where $\bx_{\perp1}$ denotes the component of $\bx$ perpendicular to $\bfh_1$. This simplicity around filament 1 comes from our choice of axes. $\phi_3$ comes from Kawasaki's theorem used on the vertex labeled $\bg_4$, together with the law of cosines on the triangle:
\begin{equation}
\cos\phi_3=\coslaw\left[(\bg_4-\bg_2)_{\perp1}, (\bg_4-\bg_3)_{\perp1}, (\bg_3-\bg_2)_{\perp1}\right],
\label{eqn:cosphi3}
\end{equation}
where $\coslaw({\bf a},{\bf b}, {\bf c})\equiv(c^2-{\bf a}\cdot{\bf b}-{\bf b}\cdot{\bf c})/(2ab)$. A similar equation gives $\phi_4$. With these, a pair of simultaneous equations containing only $\theta_2$ and $\theta_3$ comes from using the law of cosines to relate $(\bfh_{1\perp3}\cdot\bfh_{2\perp3})$ and $(\bfh_{1\perp2}\cdot\bfh_{3\perp2})$ to $\bg_i$. A similar pair of equations gives $\theta_2$ and $\theta_4$. We numerically solved these two pairs of equations with Mathematica, which did not find closed-form results. In all cases tried, single solutions were found, and the $\theta_2$'s found in both pairs of simultaneous equations agreed.

\begin{figure}
  \begin{minipage}[b]{0.495\linewidth}
    \begin{center}
     \includegraphics[width=1\columnwidth]{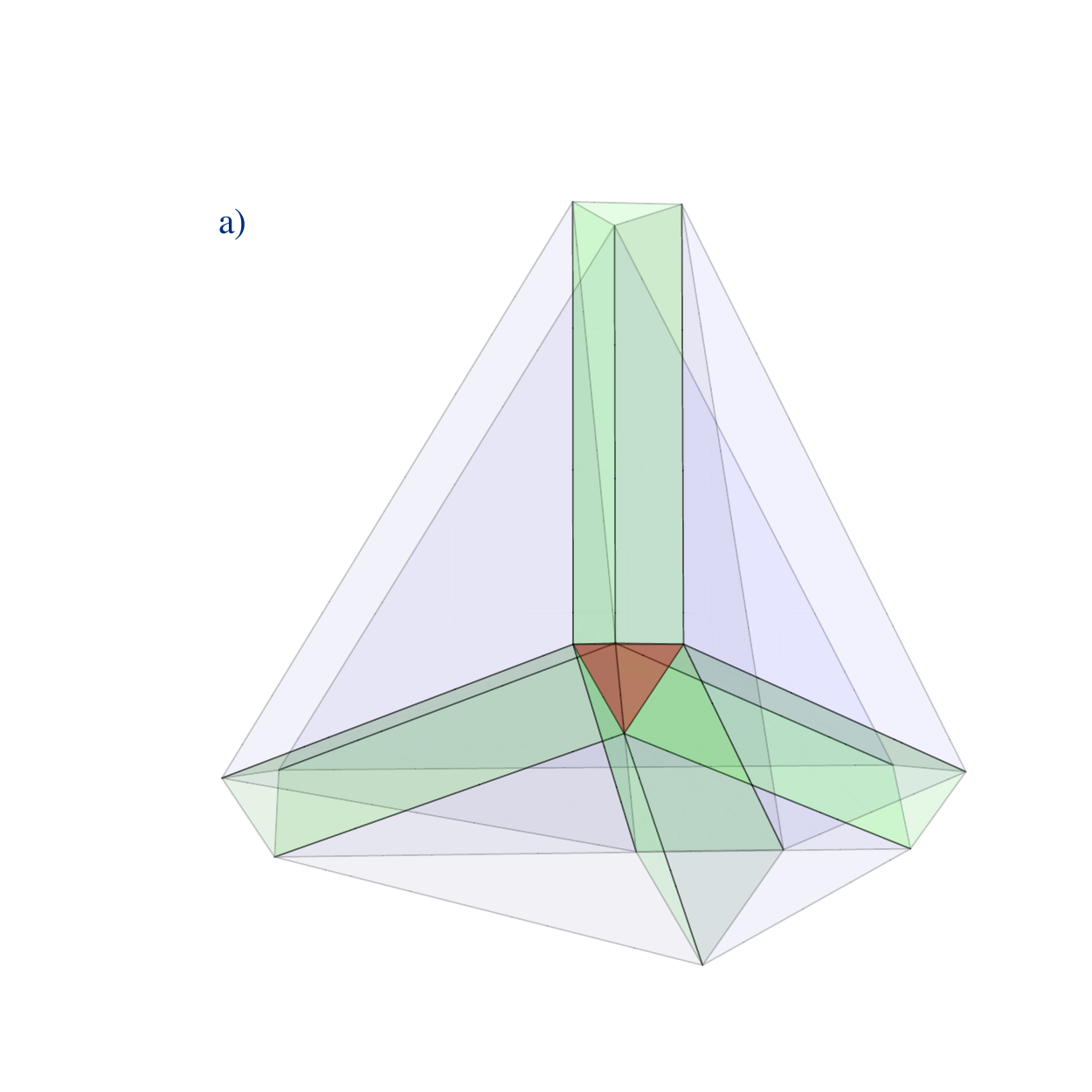}
     \includegraphics[width=1\columnwidth]{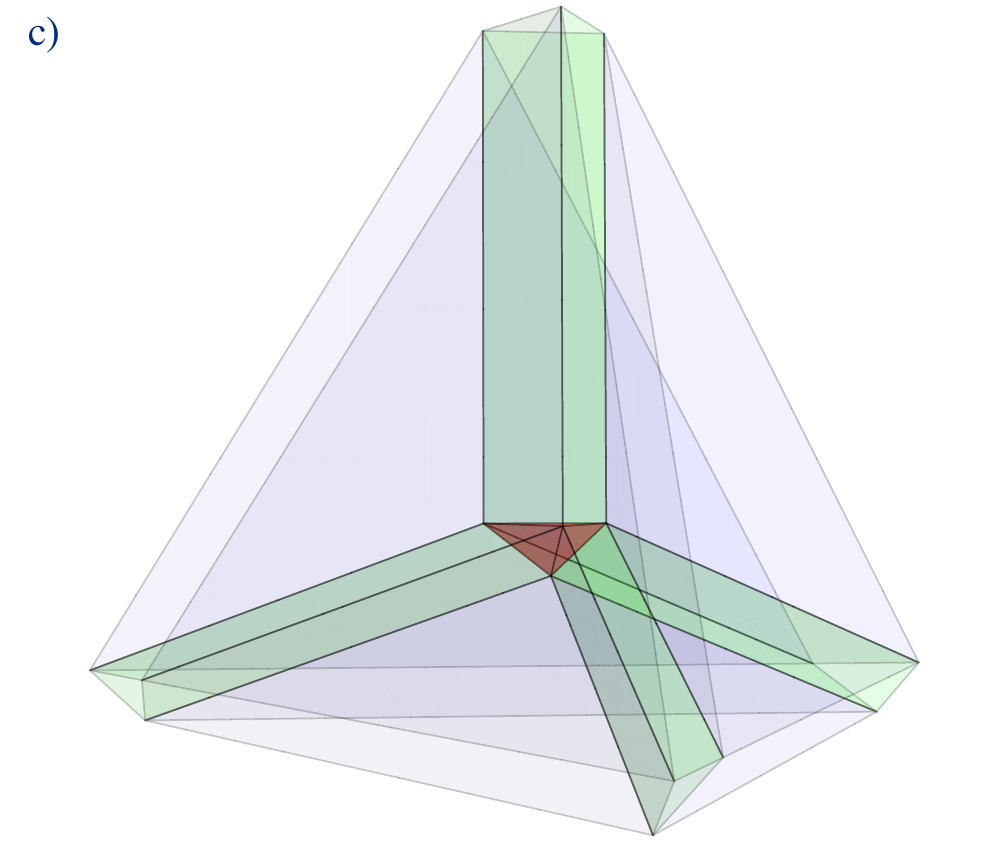}
    \end{center}  
  \end{minipage}
    \begin{minipage}[b]{0.495\linewidth}
    \begin{center}
     \includegraphics[width=1\columnwidth]{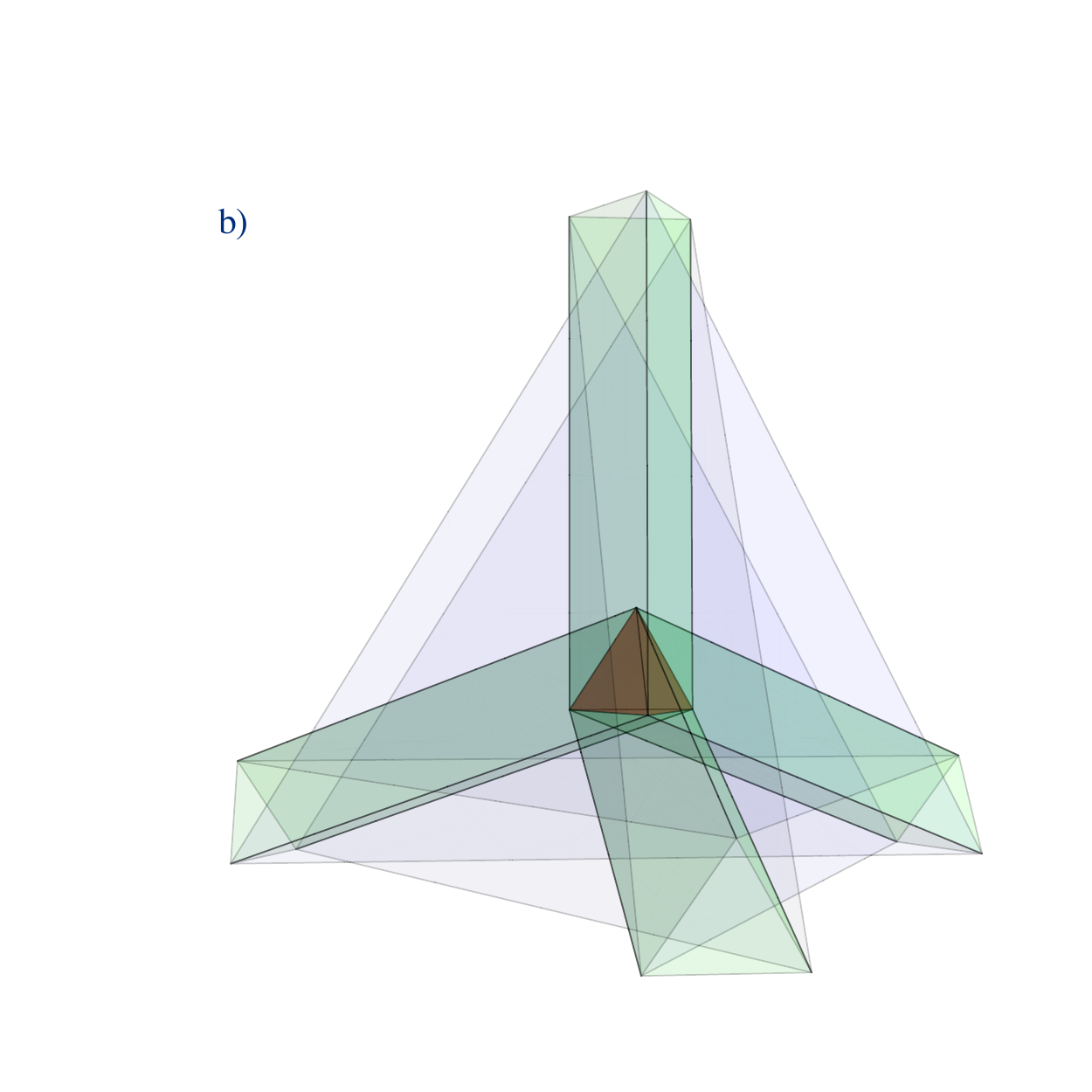}
     \includegraphics[width=1\columnwidth]{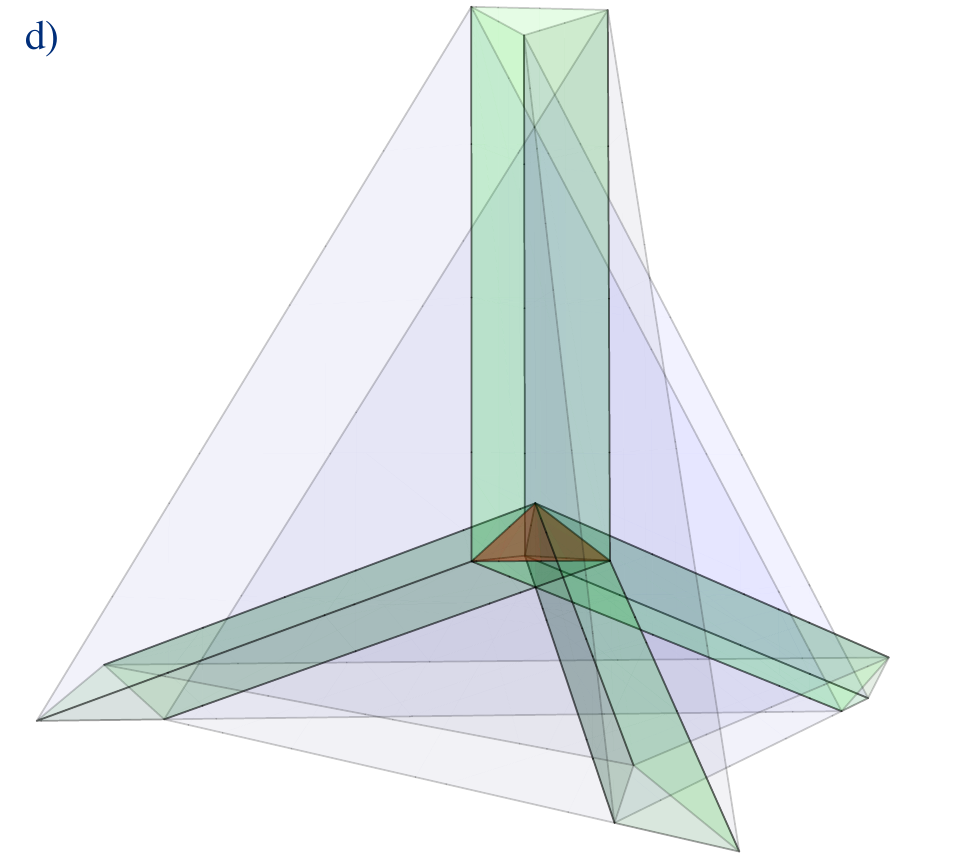}
    \end{center}  
  \end{minipage}
  \caption{Tetrahedral-collapse models/tetrahedral twist folds. Filament creases are indicated by triangular tubes, intersecting at the central node. Wall creases extend from filament edges through the thin lines drawn between filaments.  {\bf Left:} Pre-folding/collapse (Lagrangian). {\bf Right:} Post-folding/collapse (Eulerian). {\bf Top}: An irrotational model ($\alpha_1=\pi/2$). Each filament vector $\bfh_i\perp$ a face of the central tetrahedron. Walls, filaments, and the node invert along their central planes, axes, and point, but remain connected as before. Void regions simply move inward. All 15 initial regions overlap at the center. {\bf Bottom}: A rotational model ($\alpha_1=\pi/6$). The top filament rotates counter-clockwise by $\pi/3$, while the smaller, bottom filaments rotate clockwise by $2\pi/3$. See \protect\url{http://skysrv.pha.jhu.edu/~neyrinck/TetCollapse} for an interactive 3D model.}
    	\label{fig:regtet}
\end{figure}

Irrotational regular-tetrahedral collapse is the simplest, most symmetric polyhedral collapse. Its filaments in Lagrangian space, each colored differently, are shown in Fig.\ \ref{fig:regtet}a. Here, each $\bfh_i$ points directly away from each $\bg_i$. It obeys Kawsasaki-3D: the solid angle subtended by a Lagrangian void region around a vertex of the tetrahedron is $\pi$ sr, since if the tetrahedron were not there, the 4 identical void regions would meet in a point, and the 4$\pi$ sr would be divided equally among the void regions. Each of the 3 filaments meeting at a vertex subtends $\pi/3$ sr. So the sum of solid angles subtended by odd-parity regions is $2\pi$. As with irrotational equilateral-triangular collapse, all streams of the crease pattern overlap at the center, so voids, walls, filaments, and nodes characteristically have 1, 3, 7 (like a 2D node), and 15 streams, respectively.

Keeping the node regular-tetrahedral, but rotating it along $\bfh_1$, produces rotations in all filaments; also, angles between filaments change from the equiangular $109.5^\circ$. By symmetry around $\bfh_1$, $\theta_3=\theta_4=\theta_2$, but we found that $\theta_2(\alpha_1)\approx \theta_2^{\rm approx}(\alpha_1)\equiv(109.5^\circ-90^\circ) \sin \alpha_1+90^\circ$, to within $0.5^\circ$. Filament 1 remains equilateral in cross-section, but the others elongate.

A particularly interesting rotational model retains $109.5^\circ$ angles between filaments, and equilateral filament cross-sections. It involves an irregular tetrahedral node, with an equilateral face opposite (and $\perp$ to) $\bfh_1$, but with height $h=h_0\sin\alpha_1$, where $h_0$ is the height if the tetrahedron were regular. A kinematic version of this model could have two parameters that vary with time: $\alpha_1$, and an overall scale. The behavior is special at $\alpha_1=\pi/6$, as shown in the bottom panels of Fig.\ \ref{fig:regtet}. Here, the top filament, of side length $s$, rotates in one way by $\pi/3$ from its initial to final state. The bottom filaments, with cross-sections of side length $s/\sqrt{3}$, rotate in the opposite way, by $2\pi/3$. The linear density ($\propto s^2$) of the top filament equals the sum of the linear densities of the three bottom filaments.  Future work will further explore this model.

We also verified Kawasaki-3D numerically in several models specified with different $\bg_i$. Kawasaki-3D takes the following form around vertex $\bg_4$, summing solid angles subtended by all regions of odd parity (filaments and voids):
\begin{equation}
\Omega(\bfh_1,\bfh_2,\bfh_3)+\Omega(\bg_2-\bg_4,\bg_3-\bg_4,\bfh_1)+\Omega(\bg_3-\bg_4,\bg_1-\bg_4,\bfh_2)+\Omega(\bg_1-\bg_4,\bg_2-\bg_4,\bfh_3)=2\pi,
\end{equation}
where $\Omega(\ba,\bb,\bc)$ gives the solid angle subtended by a triangle with vertices along vectors $\ba$, $\bb$, and $\bc$ \cite{sphangle}:
\begin{equation}
\Omega(\ba,\bb,\bc)=2 \tan^{-1}\frac{|\ba\cdot(\bb\times\bc)|}{abc+(\ba\cdot\bb)c+(\ba\cdot\bc)b+(\bb\cdot\bc)a}.
\end{equation}
A Mathematica notebook which solves for the five $\bfh_i$ angles in terms of $\bg_i$ is available from the author upon request.

\subsection{Physical Quantities}
It is helpful to explicitly connect quantities in polyhedral collapse with physical quantities that one might measure, e.g.\ in a cosmological simulation. The {\it mass} of a node (or halo, or galaxy) is simply its volume $V$ in $\mL$ (times a constant density $\rho_0$). Note that other streams (for tetrahedral collapse, up to 15, the total number of streams) will overlap with the node once it is folded in $\mE$; this could be added to the mass, as well. Similarly, the mass per unit length of a filament is its cross-sectional area $A$ times $\rho_0$.  An `angular momentum' can be defined for a node, $L=2\alpha_{\rm node}V$, where $2\alpha_{\rm node}$ is the angle by which it turns during collapse. For a filament, an angular momentum per unit length $L/\ell=2\alpha A$. In further work, we plan to give relations between these quantities in polyhedral-collapse models.
 
\section{Discussion}
This origami approximation in large-scale structure, that the dark-matter sheet does not stretch, gives an idealized cosmic web, with convex voids and nodes. The no-stretch condition is generally false in reality, but it may hold closely enough in some regimes to give useful results. In this approximation, the convexity of the various shapes, as well as the inability to form a structure in isolation, gives a simple topology: voids, as delineated by multistream regions, are convex and do not percolate through walls. Voids in `reality' (full cosmological $N$-body simulations) seem roughly convex only if using a density-based criterion to define them, but not by delineating them with multistream regions \cite{NeyrinckEtal2013,FalckNeyrinck2015}. (Note that density-based and multistreaming-based criteria exactly coincide in this origami approximation, where density can increase at an Eulerian point only by increasing the number of streams.) This is also qualitatively like a model in which voids are Voronoi cells \cite{IckeVdW1991}.

Here we began to study the purely geometric `polyhedral collapse' model. The arrangement of filaments (one per face) around a polyhedral node determines the physical properties such as rotation of all of the pieces in the final conditions.  This model could substantially aid the understanding of how filaments and galaxies rotate (or do not) based on their spatial arrangement, a field of much study \cite{AragonEtal2007,CodisEtal2012,AragonYang2013,DuboisEtal2014}. There is even observational evidence for a correlation between galaxy spins and their arrangement in filaments and walls \cite{TempelLibeskind2013}.  This would give an `intrinsic alignment' between the observed ellipticities of nearby galaxies \cite{CodisEtal2015}, a systematic effect that must be overcome for weak lensing measurements to realize their full potential as a key test of cosmological questions, such as why the Universe seems to be accelerating in its expansion.  

Many of the results given here, particularly in 3D, were numerical. This is fine for comparison to cosmological simulations, as we plan to do. But there is much room for further rigorous mathematical study of polyhedral collapse, both of isolated nodes, and of how networks of collapsed polyhedra behave together. For a more astronomical discussion of the origami approximation, including some explicit velocity fields, see \cite{Neyrinck2014}.

\bibliographystyle{akpbib}
\bibliography{refs}

\end{document}